\documentclass[adp,fleqn]{w-art}
\usepackage{graphicx}
\chardef\bslash=`\\ 

\newcommand{\hatd}[2]{\hat{ #1 }^{\dagger}_{ #2 }}
\newcommand{\hatn}[2]{\hat{ #1 }^{\ }_{ #2 }}

\newcommand{\eqsa}[1]{\begin{eqnarray} #1 \end{eqnarray}}

\begin{document}
\DOIsuffix{theDOIsuffix}
\pagespan{1}{}
\Dateposted{1 April, 2011}




\title[Theory of pseudogap in doped Mott insulators]{Theory of pseudogap and superconductivity \\ in doped Mott insulators}


\author[M.\ Imada]{Masatoshi Imada\inst{1}} 
\address[\inst{1}]{Department of Applied Physics, University of Tokyo, Hongo, Tokyo, 113-8656, Japan}
\author[Y. Yamaji]{Youhei Yamaji\inst{1,2}
}
\address[\inst{2}]{Department of Physics, Rutgers University,
Piscataway, New Jersey 08854, USA}
\author[S. Sakai]{Shiro Sakai\inst{3}
}
\address[\inst{3}]{Institute for Solid State Physics, Vienna University of Technology, 1040 Vienna, Austria}
\author[Y. Motome]{Yukitoshi Motome\inst{1}}

\begin{abstract}
  Underdoped Mott insulators provide us with a challenge of many-body physics.  Recent renewed understanding is discussed in terms of the evolution of pole and {\it zero} structure of the single-particle Green's function. Pseudogap as well as Fermi arc/pocket structure in the underdoped cuprates is well reproduced from the recent cluster extension of the dynamical mean-field theory. Emergent coexisting zeros and poles set the underdoped Mott insulator apart from the Fermi liquid, separated by topological transitions.  Cofermion proposed as a generalization of exciton in the slave-boson framework  accounts for the origin of the {\it zero} surface formation. The cofermion-quasiparticle hybridization gap offers a natural understanding of the pseudogap and various unusual Mottness.  Furthermore the cofermion offers a novel pairing mechanism, where the cofermion has two roles: It reinforces the Cooper pair as a pair partner of the quasiparticle and acts as a glue as well.  It provides a strong insight for solving the puzzle found in the dichotomy of the gap structure.     
\end{abstract}
\maketitle                   





\section{Introduction}\label{Sec1}
\label{sect1}
In the history of the cuprate superconductors, the pseudogap was first
experimentally identified as a spin gap in the relaxation time $T_1$ of
the nuclear magnetic resonance\cite{Yasuoka} and in the subsequent Knight shift measurement\cite{Takigawa}. 
Later on, the whole gap structures were found in other probes including
charge channels such as the optical conductivity\cite{Rotter,Homes_Timusk} and the angle resolved 
photoemission spectroscopy (ARPES)\cite{Shen} as well as in thermodynamic probes such as the 
specific heat\cite{Loram}. 
Theoretically, there exist a number of interpretations and proposals attempting to account for the origin of the pseudogap, while it is still under active debates\cite{RMPImada}.
The mechanism of the pseudogap formation is widely believed to be intimately correlated with the mechanism of the high temperature superconductivity itself. 

In 1989, Metzner and Vollhardt\cite{Vollhardt} pioneered the dynamical mean-field
theory (DMFT). Recent cluster extensions of DMFT\cite{Maier} have enabled to numerically answer the question of gap and pseudogap formations by a systematic increase of the cluster size\cite{Kotliar,SakaiPRL,SakaiPhysica,Sakai2010}. 
In this short article, we propose a physical intuition to understand essence of the pseudogap formation and the mechanism of the high-$T_c$ superconductivity, after reviewing recent numerical results on the 2D Hubbard model achieved by the cluster DMFT followed by a consistent description attained in the improved slave-boson formalism called the cofermion theory\cite{YamajiPRL,Yamaji2011}.

A trivial origin of the gap formation in the electronic spectra in condensed matter is found in the periodic potential of the nuclei in crystals.  Aside from such external origins ascribed to the translational symmetry breaking of lattice coupled to electronic periodical density modulations, we have various intrinsic and electronic origins of the gap formation around the Fermi level in condensed matter systems. In most of them, however, a gap in electronic spectra around the Fermi level can be understood from their hybridizations with some other modes in the mean-field picture.   
Suppose the Hamiltonian is given by the single-particle form as 
\begin{equation}
H = \sum_{i} \xi_{\alpha_i} \hatd{c}{\alpha_i}\hatn{c}{\alpha_i} + \sum_{i} \epsilon_{\beta_i} \hatd{d}{\beta_i}\hatn{d}{\beta_i} + \sum_{i}\Delta_i\hatd{c}{\alpha_i}\hatn{d}{\beta_i}+{\rm H.c.},
\label{MFE}
\end{equation}
where the quasiparticle with the creation (annihilation) operator $\hatd{c}{\alpha_i}$ ($\hatn{c}{\alpha_i}$) 
hybridizes with the particle $\hatd{d}{\beta_i}$ ($\hatn{d}{\beta_i}$) with the quantum numbers $\alpha_i$ and $\beta_i$, respectively. 
Then a hybridization gap proportional to $\Delta_i$ opens in the single-particle spectra after the diagonalization.

For instance, in the mean-field theory, the antiferromagnetic gap is caused by the hybridization of electrons at a wave number $q$ with the electrons themselves but at a different wavenumber $q+Q_0$, where $Q_0$ is the ordering vector.
A $\uparrow$-spin electron with momentum $k$
hybridizes with a $\downarrow$-spin electrons with momentum $k+Q_{0}$,
as the term $\Delta_{\rm AFM}[\hatd{c}{k\uparrow}\hatn{c}{k+Q_{0}\downarrow}+{\rm H.c.}]$,
in the presence of the mean field $\Delta_{\rm AFM}$.
As a result,
the self-energy for the $\uparrow$-spin electron diverges
at the momentum $k$ where the $\downarrow$-spin electron with momentum $k+Q_{0}$ has a pole.

When
the bare dispersion of $\hatd{c}{k\sigma}$ in the absence of $\Delta_{\rm AFM}$ is given by $\xi_k$,
the Green's function of 
the electron $\hatd{c}{k\sigma}$
that
hybridizes with the electron $\hatd{c}{k+Q_{0}\overline{\sigma}}$
and its self-energy are given
as
\eqsa{
	G(k,\omega)
	&=&
	\frac
	{1}
	{\displaystyle\omega-\xi_{k}-\Sigma(k,\omega)},
	\ \ \  {\rm and} \ \ \ 
	\Sigma(k,\omega)
	=
	\frac{\Delta_{{\rm AFM}}^{2}}{\omega-\xi_{k+Q_{0}}}.
	\label{AFM_zero}}
This is also similar to the case of the pairing interaction in BCS theory, although the ``hybridization" in the  particle-hole excitation is replaced with the particle-particle channels.

With the symmetry breaking, a zero surface appears in the single-particle Green's function.
The zero surface is determined from the self-energy pole, where $\Sigma=\pm \infty$ results in the breakdown of the perturbation expansion.  The pole of the self-energy is nothing but the pole of the partner of the hybridization, namely, the electron at $k+Q_0$ for the above antiferromagnetic order.  

\begin{figure}[htb]
\begin{minipage}[t]{.6\textwidth}
\includegraphics[width=0.55\textwidth]{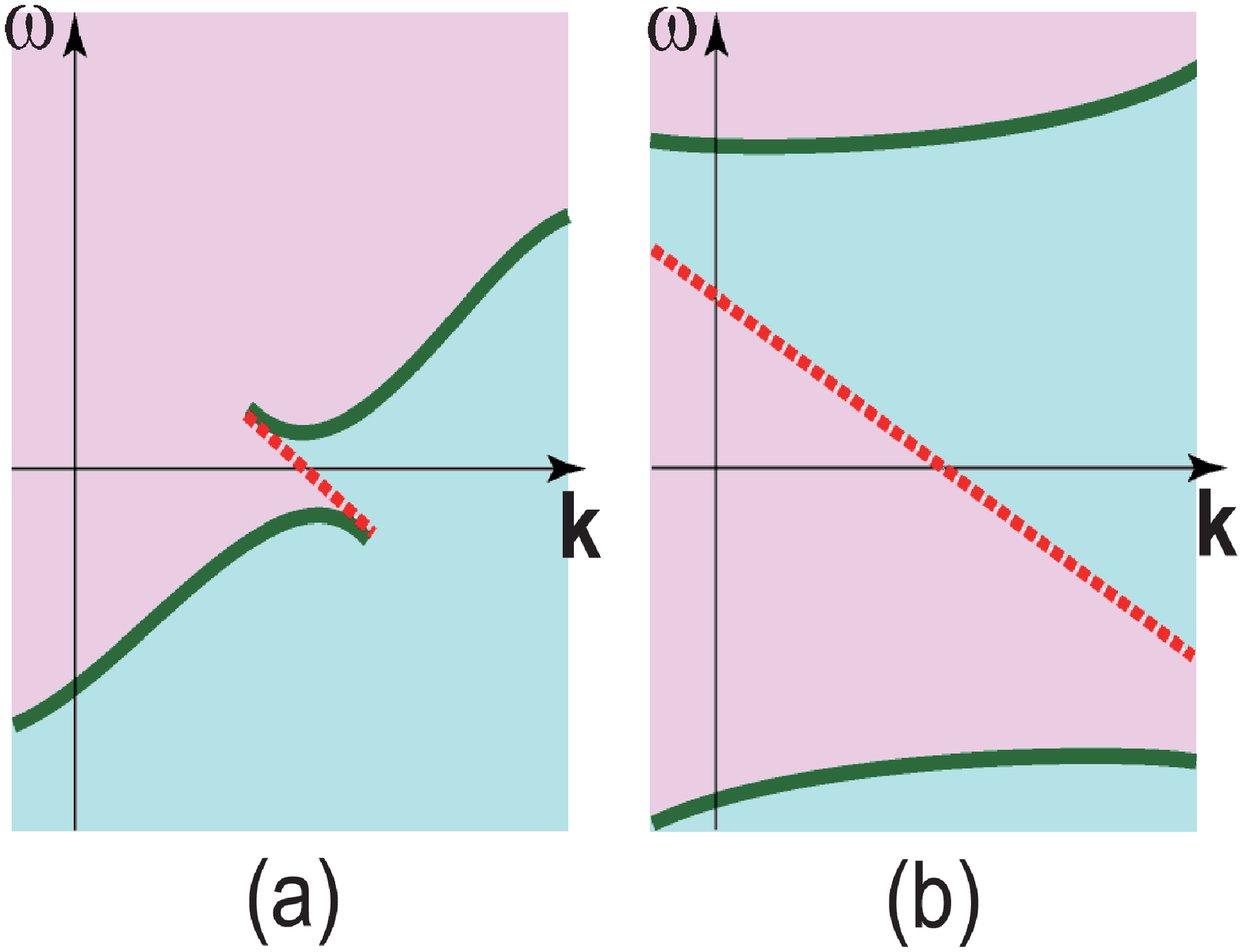}
\caption{Schematic structure of pole surface (bold (green) solid curve) and zero surface (bold (red) dashed curve). (a): A small gap opening at the hot spot just splits the pole surface by creating the zero surface bridging the pole surfaces. (b): Gap and pole-zero structure in large gap region. Positive Re$G$ region (pink) is distinguished from negative region (light blue). Only by the combination of the pole and zero, a consistent description of the sign change in ${\rm Re} G$ is attained.}
\label{ZeroSchematic}
\end{minipage}
\hfil
\begin{minipage}[t]{.35\textwidth}
\includegraphics[width=0.77\textwidth]{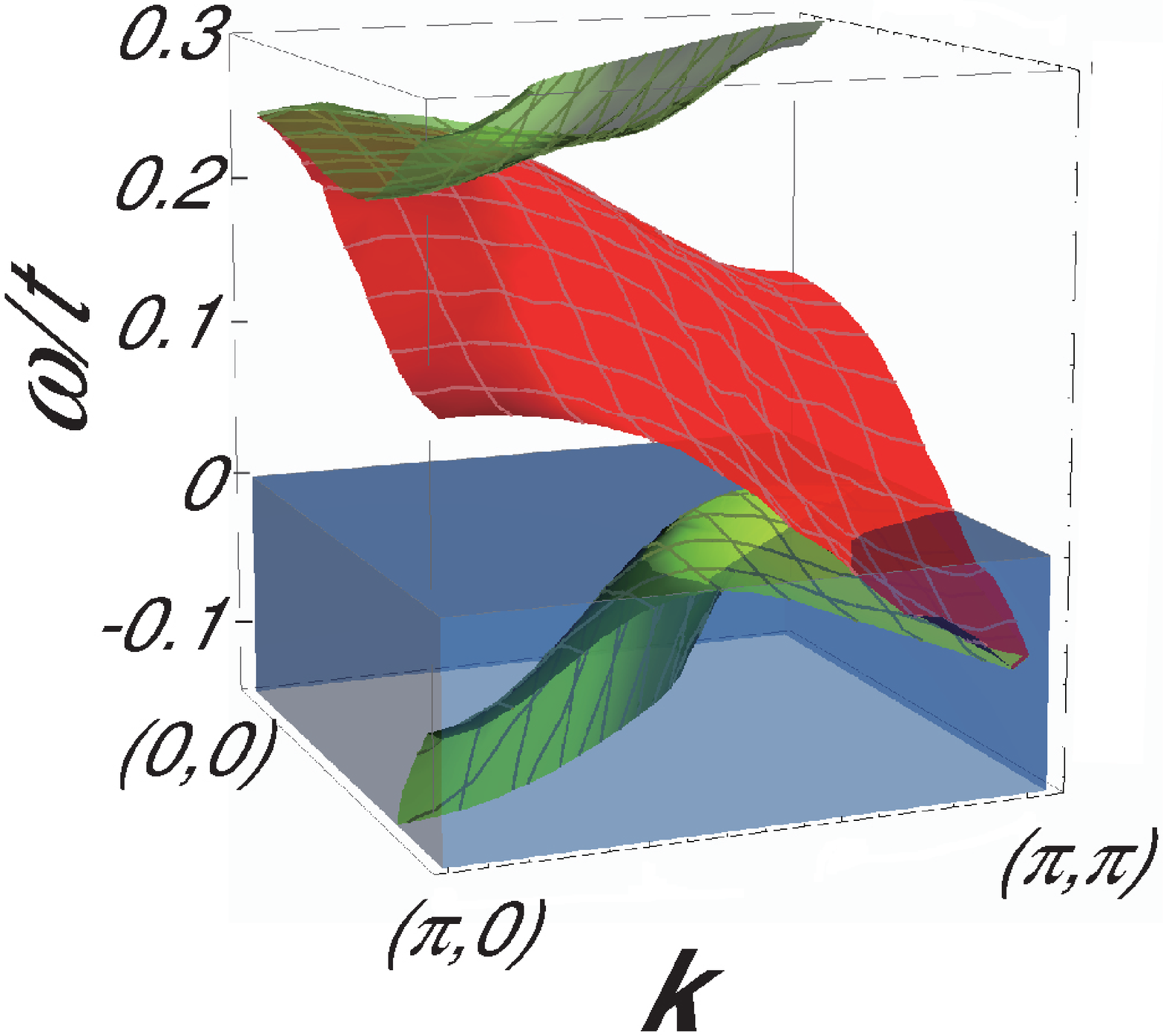}
\caption{Poles (green) and zeros (red) of single-particle Green's function in momentum $k$ and energy $\omega$ space for Hubbard model on a square lattice at $U/t=8$ and doping concentration $\delta =0.09$ revealed by  cDMFT calculation\cite{SakaiPRL}. Occupied regions for an electron are filled with ``aqua".}
\label{ZeroCDMFT}
\end{minipage}
\end{figure}

Now the gap formation and the disappearance of the Fermi surface leading to the insulator 
are caused by the emergence of the zero surface which splits the original single dispersion into two. In other words, thus formed two dispersions are bridged by the zero surface in the gap region as we see in Fig.\ref{ZeroSchematic}, which consistently explains how the positive definite ${\rm Re} G$ in the largely positive $\omega$ region changes its sign into the negative definite one in the largely negative $\omega$.      

The gap formation and the disappearance of the Fermi surface are thus understood in a general framework for symmetry broken phases together with the  emergence of the zero surface.  A question, however,  arises how the Mott gap as well as the cuprate pseudogap in the absence of the symmetry breaking should be understood.
We propose a consistent novel picture that
answers this question. In \S \ref{Sec2} we summarize recent numerical results that are consistent with experimental results in the cuprates. In \S  \ref{Sec3}, we introduce cofermion concept as a generalization of the exciton physics and show its consistency with the numerical results in \S \ref{Sec2}. \S \ref{Sec4} describes the superconducting mechanism in the cofermion theory.  \S  \ref{Sec5} is devoted to conclusive remarks.

\section{Pseudogap in numerical results}\label{Sec2}

We first start from describing what the cluster DMFT, more specifically the cellular DMFT (cDMFT), tells us.
Details of the cDMFT method is given in the literature\cite{Kotliar,Sakai2010}.  In the results of the cDMFT calculation performed on the 2D Hubbard model defined by
\begin{equation}
H = -\sum_{\langle ij \rangle \sigma} t_{ij} \hatd{c}{i\sigma}\hatn{c}{j\sigma} +{\rm H.c.} + \sum_{i} U \hatn{n}{i\uparrow}\hatn{n}{i\downarrow},
\label{Hubbard}
\end{equation}
the zero surface indeed appears at large onsite interaction $U$. It separates the upper and lower Hubbard bands and the zero surface runs in the resultant Mott gap similarly to the symmetry broken cases as we see a schematic structure in Fig.\ref{ZeroSchematic}.

The physical origin of the zero surface in the weak coupling picture may be ascribed to the Umklapp scattering\cite{Rice}, where the Hubbard $U$ term may be decoupled as the Umklapp scattering of two electrons in the momentum space as 
\begin{equation}
U \hatd{c}{q\sigma}\hatn{c}{q+G}\langle \hatd{c}{p\sigma}\hatn{c}{p+G'}\rangle,
\label{Umklapp}
\end{equation}
where $G$ and $G'$ are reciprocal lattice vectors.
It may generate a gap first at the so-called hot spot defined as the points of the Fermi surface intersecting the border of the halved Brillouin zone. It leads to the splitting of the pole surface by the zero surface as in Fig.\ref{ZeroSchematic}(a).
In the mean-field picture given by Eq.(\ref{MFE}), the partner of the hybridization is always interpreted by the initial and final states of the Umklapp scattering. It first generates a gap only in a small portion of the Fermi surface near the hot spot. Of course this weak coupling picture may be substantially modified in the strong coupling region, where the gap opens along the entire Fermi surface as in Fig.\ref{ZeroSchematic}(b).

The zero surface may be created either by annihilating the Fermi surface as in the above Umklapp process or by simultaneous creation of a pole surface.
To make the sign of the Green's function consistent, even number of the creation or annihilation of the poles or zeros should occur. 

When a zero surface is close by, the quasiparticle band dispersion is seriously affected, particularly at the gap edge. The zero surface indeed forces the back bending of the quasiparticle dispersion near the gap edge (Fig.\ref{ZeroSchematic}(a)).  At large $U$, the zero surface with the width of the original dispersion of the hybridizing particle runs only in a middle part of the gap extended over the entire Brillouin zone by keeping basically the $U$ independent ``bandwidth" as we see in Fig. \ref{ZeroSchematic}(b).  In addition,  the ``quasiparticle" dispersions (upper and lower Hubbard bands) retain a shape of the dispersion similar to that of the original bare particle, but are renormalized to flatter and flatter ones with increasing $U$. This structure is basically similar to the symmetry broken case and understood from qualitative dependence in Eq.(\ref{AFM_zero}).

At $\omega=0$ in the Mott insulating phase, the Fermi surface  that characterizes the Fermi liquid in the conventional metallic phase is missing and there may exist only the zero surface instead of the Fermi surface.  
Since the transition between the Mott insulator and the Fermi liquid in the heavily doped region is continuous in these calculations, one has to have a region that is characterized by the coexistence of the poles and zeros at $\omega =0$ in between the two phases.  

This is indeed the case of lightly doped Mott insulators relevant to the pseudogap region in the cuprates. The cDMFT with the solvers of the exact diagonalization at temperature $T=0$ and the continuous-time quantum Monte Carlo method at $T>0$  for the available clusters up to 8 sites with the cumulant periodization in the Brillouin zone consistently show the coexistence (see Fig.\ref{ZeroCDMFT}\cite{Kotliar,SakaiPRL,Sakai2010}). Upon doping of small concentration of holes into the Mott insulator, the Fermi level moves to the top of the lower Hubbard band. For large $U/t$, the backbone Mott gap structure with a zero surface similar to Fig.\ref{ZeroSchematic}(b) is still retained at a high energy, while  Fig.\ref{ZeroCDMFT} shows that an intricate small-energy structure emerges hierarchically near the Fermi level characterized by a low-energy zero surface sandwiched by the two pole surfaces of quasiparticles.  This second zero surface at low energy generates the pseudogap, leaving a hole pocket of Fermi surface in the bottom of the pseudogap.  The pseudogap has a $d$-wave-like structure below the Fermi level and well accounts for the ARPES results.  

However, if we consider the total structure including the part above the Fermi level, it in fact has the structure of a full gap with the isotropic $s$-wave like symmetry (see Fig.\ref{ZeroCDMFT}), in contrast to the expectation by the $d$-wave picture\cite{Tohyama}. 
The unoccupied part above the Fermi level has not been well probed so far because of the experimental difficulty as in the poor resolution in the inverse photoemission. The pseudogap structure critically poses restrictions to the theories as well as to the mechanism of the superconductivity. The present theory and proposal can be critically tested by measuring whether the full-gap structure indeed exists. It is desired in future high-resolution experiments probing the unoccupied part.

Aside from the unexplored and unoccupied spectra, a number of unconventional features of the high-$T_{\rm c}$ cuprates in the underdoped region are also nicely reproduced by the cDMFT results, where this coexistence of the zero and the Fermi surface holds the key\cite{Sakai2010}. They include not only the arc/pocket formation and pseudogap, but also the kink structure, back-bending dispersion, asymmetry of the density of states, and so on.

In addition, the region of coexisting pole and zero surfaces cannot be connected adiabatically neither with the Fermi liquid in the overdoped region nor with the Mott insulator at zero doping. Therefore, this coexisting region is separated not by any symmetry breaking transition but by the topological transitions of the Fermi surfaces.  A number of non-Fermi-liquid properties in the underdoped region are characteristic of this topologically distinct phase. A typical topological distinction and their transitions are illustrated schematically in Fig.\ref{NonFermiLiquid}.  

\section{Cofermion}\label{Sec3}

The pseudogap formation has firmly been established in the numerical results on the doped Mott insulator, while it is useful to understand the intuitive mechanism.
As the origin in the cuprates, there are several proposals reducing to symmetry breakings or their precursors.
For instance, antiferromagnetic or charge order, $d$-density wave, and superconductivity itself and their precursory fluctuations have been proposed.  
In this article, since the apparent symmetry breaking is not necessarily clear in the cuprates,  we instead propose a natural mechanism of the pseudogap formation generic but inherent in lightly doped Mott insulators, by extending the concept of excitons, which  has played a significant role in understanding excitations of semiconductors.     

\begin{figure}[htb]
\begin{minipage}[t]{.62\textwidth}
\includegraphics[width=.65\textwidth]{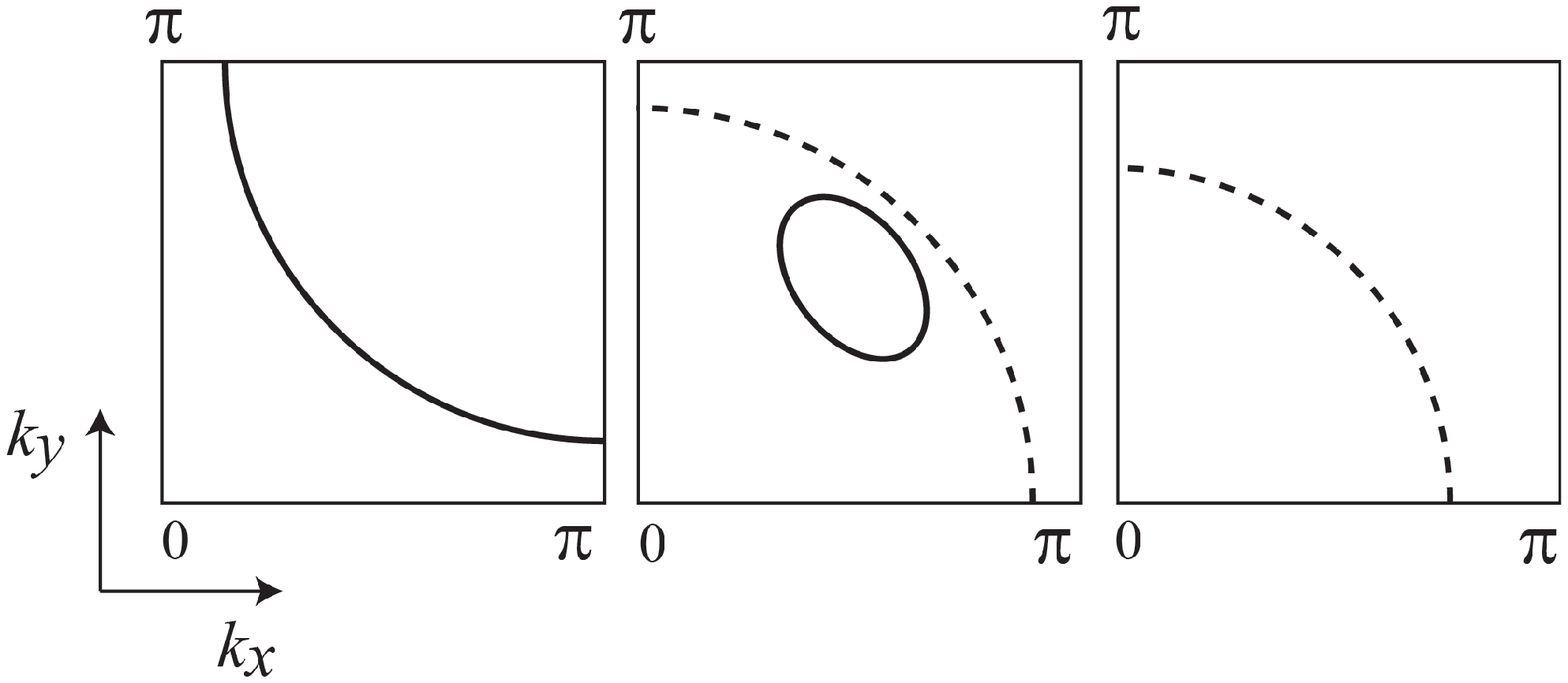}
\caption{Schematic illustration of three phases in the 2D Brillouin zone. Left:  Overdoped Fermi liquid phase with only the pole surface(solid curve). Middle: Possible realization of non-Fermi-liquid with coexisting pole and zero (dashed curve) surfaces in the underdoped region. Right: Mott insulator with only the zero surface (dashed curve).  Three phases are separated by topological transitions. Between the left and the middle, at least two topological transitions (the zero-pole pair creation and an additional Lifshitz transition of the pole surface) occur to bridge these two phases.}
\label{NonFermiLiquid}
\end{minipage}
\hfil
\begin{minipage}[t]{.35\textwidth}
\includegraphics[width=.67\textwidth]{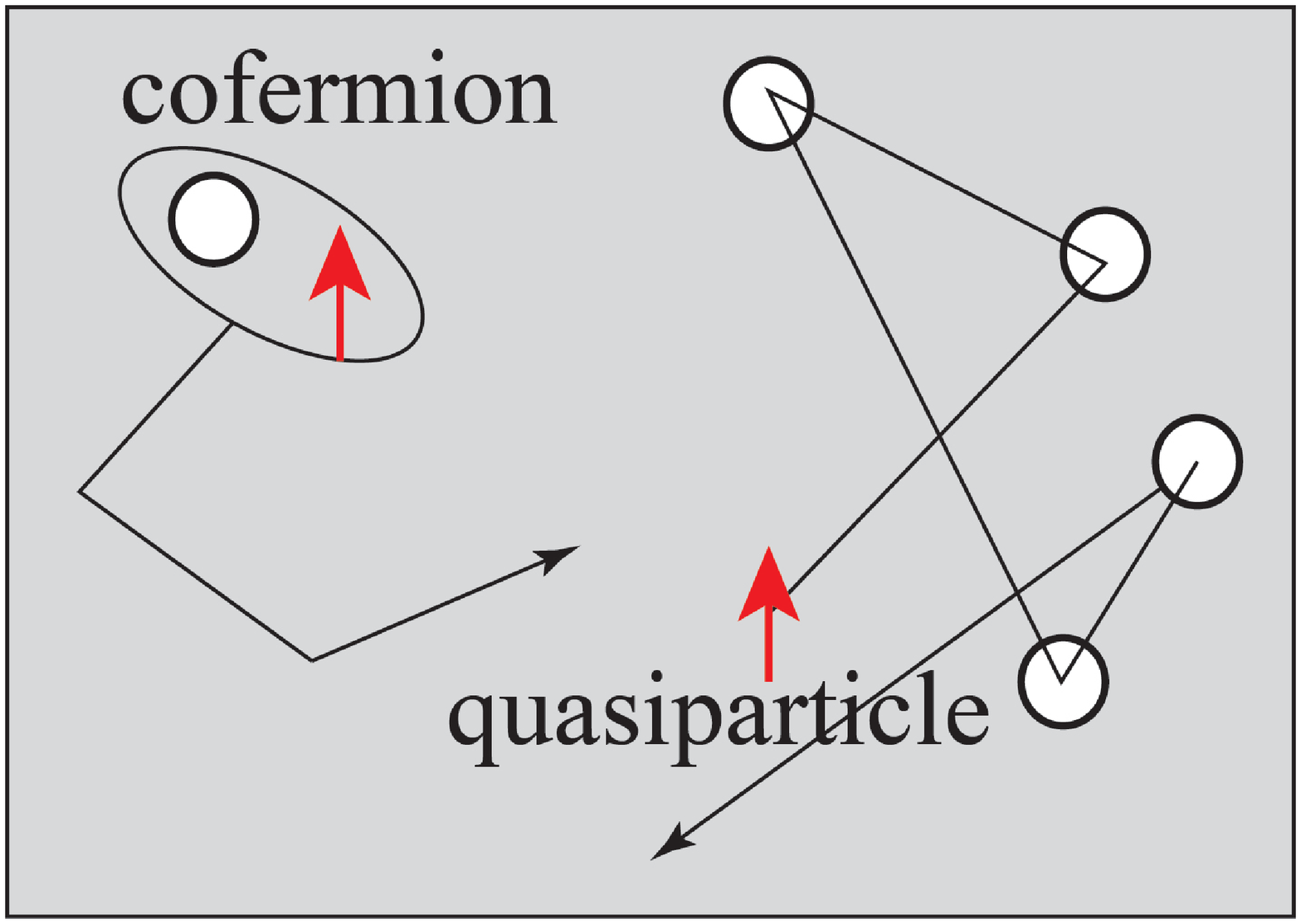}
\caption{Schematic illustration of cofermion and quasiparticle in the sea of half-filled electrons depicted as gray background.
These two excitations both have 1/2-spins. They also hop and becomes extended in the real space giving their dispersions. White circles represent holes and (red) bold arrows represent added electrons.}
\label{cofermion_schematic}
\end{minipage}\end{figure}
In the lightly hole doped region, as we mentioned above, the Mott gap is still a robust underlying structure, where a similar zero surface persists, while a low-energy-scale structure near the gap edge also evolves hierarchically on top of it. 
Let us consider the electronic excitation in this circumstance: There are three ways of adding electrons to the unoccupied part ($\omega >0$).  
One is the high-energy excitation that adds an electron to a singly occupied site generating a doubly occupied site counted in the upper Hubbard band.  Other two ways are restricted within the low-energy structure near the Fermi level, namely, quasiparticle and cofermion excitations as are illustrated in Fig.\ref{cofermion_schematic}: Quasiparticles have a character of spatially extended state that mainly hop through empty (holon) sites.
On the other hand one may add an electron near a holon site by forming a bound state of an added electron and the existing holon as in the case of excitons. Such an exciton is a tightly bound state with the binding energy of the Mott gap in the Mott insulating state. In fact, this exciton-type excitation is the only one allowed in the large $U$ limit, where it is localized and a classical object.  Even at large but finite $U$ in the lightly doped systems, this excitonic excitation may clearly be distinguished from quasiparticles when the added electron forms a state bound with the holon as one naturally expects.  

The binding energy is, however, weakened because of the screening of the attractive force of the exciton by other doped holons and the bound state becomes extended over several sites.  In fact, the exciton binding energy proportional to $U$ in the Mott insulating state is roughly screened in the RPA to
$W=U/(1-UP)$, where $P$ is the polarizability.  Here, $P$ is essentially given by the sum of the part proportional to the doping concentration $\delta$ as $\sim -a\delta$ and the polarization $P_0$ in the Mott insulator, where $a$ is a positive constant. Since the polarization $P_0$ is basically proportional to $1/U$ for large $U$, by taking $P_0=b/U$, the screened binding energy is given by 
\eqsa{
	W
    &=&
    \frac{1}{a\delta/t+(b+1)/U}
	\label{W}
}
in the underdoped region with constants $a$ and $b$.
We call this bound state a {\it cofermion}\cite{YamajiPRL}. Equation (\ref{W}) indicates that the binding energy of the cofermion quickly reduces from the order of the Mott gap $U$ to the order of $t$ at the doping concentration $\delta \sim 1/a$, where $a$ may be roughly the order of 10 in the realistic condition of the cuprates.  Such a quick reduction of the binding energy is evidenced in the optical conductivity\cite{Takagi}. The quick shift and evolution of the mid infrared peak in the optical conductivity from the charge transfer gap to the Drude peak in the doping process from the Mott insulator to a metal may well be accounted for by the unbinding transition of a cofermion to a quasiparticle and a hole.    

The coherent weights of the cofermion 
and quasiparticle in the unoccupied part ($\omega>0$) 
 are scaled by the free holon density $\delta$ and vanish
in the Mott insulator, because they can be generated only around a free holon site. Nevertheless, the incoherent cofermion still remains in the occupied part even in the Mott insulator.  

This cofermion may propagate with an effective transfer $t_{cf}$, although it does not have a charge, same as the conventional exciton. In the cofermion theory, in contrast to the conventional exciton, the holon follows Bose statistics and thus the cofermion is regarded as a fermion.  The hopping of the cofermion takes place in the second order process in the strong coupling expansion (namely in the expansion in terms of $t/W$) as $t_{cf}\sim t^2/W$, where, in the intermediate state of the second order perturbation, the bound state in a cofermion is dissolved virtually.  It is reduced to the spin exchange by the superexchange coupling $J$ in the Mott insulating state, where the cofermion dispersion may merge into the spin exchange or alternatively a ``spinon" dispersion 
in the occupied part. For nonzero $\delta$, this neutral excitation evolves to a cofermion and its hopping amplitude $t_{cf}$ is basically scaled by $t_{cf}\sim a\delta t$ (see Eq.(\ref{W})), when $a\delta/t$ exceeds $(b+1)/U$.
 
\begin{figure}[htb]
\begin{minipage}[t]{.53\textwidth}
\includegraphics[width=1.2\textwidth]{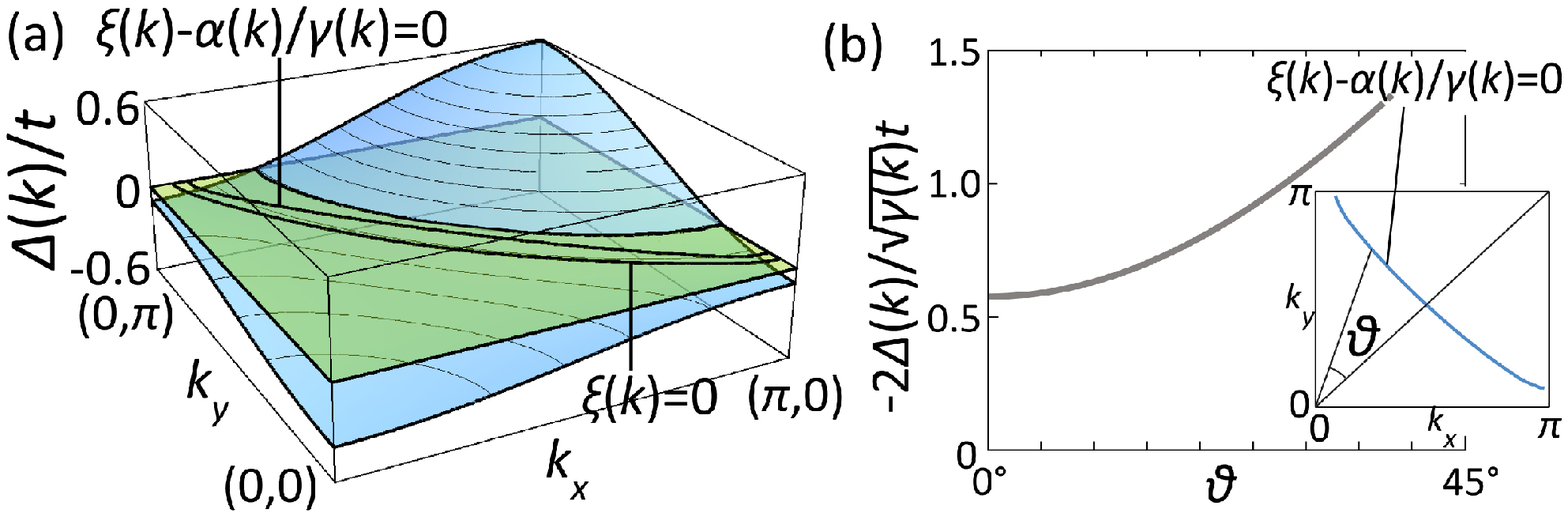}
\caption{(a) Hybridization amplitude $\Delta(k)$ (shaded curved (blue) surface) obtained from cofermion theory for 2D Hubbard model with (next) nearest-neighbor transfer $t$ ($t'$) at $t'/t=-0.25$, $U/t= 12$ and $\delta= 0.05$\cite{YamajiPRL}. Along the curve $\xi(k)-\alpha(k)/\gamma(k)=0$, $\Delta$ is always below zero (lightly shaded (green) plane) and the gap in $\Xi$ given by $2\Delta(k)/\sqrt{\gamma(k)}$ defines the pseudogap indicating a full gap structure as in (b)   
in qualitative agreement with the cDMFT result\cite{SakaiPRL}.  }
\label{Delta(k)}
\end{minipage}
\hfil
\begin{minipage}[t]{.41\textwidth}
\begin{flushright}
\includegraphics[width=.8\textwidth]{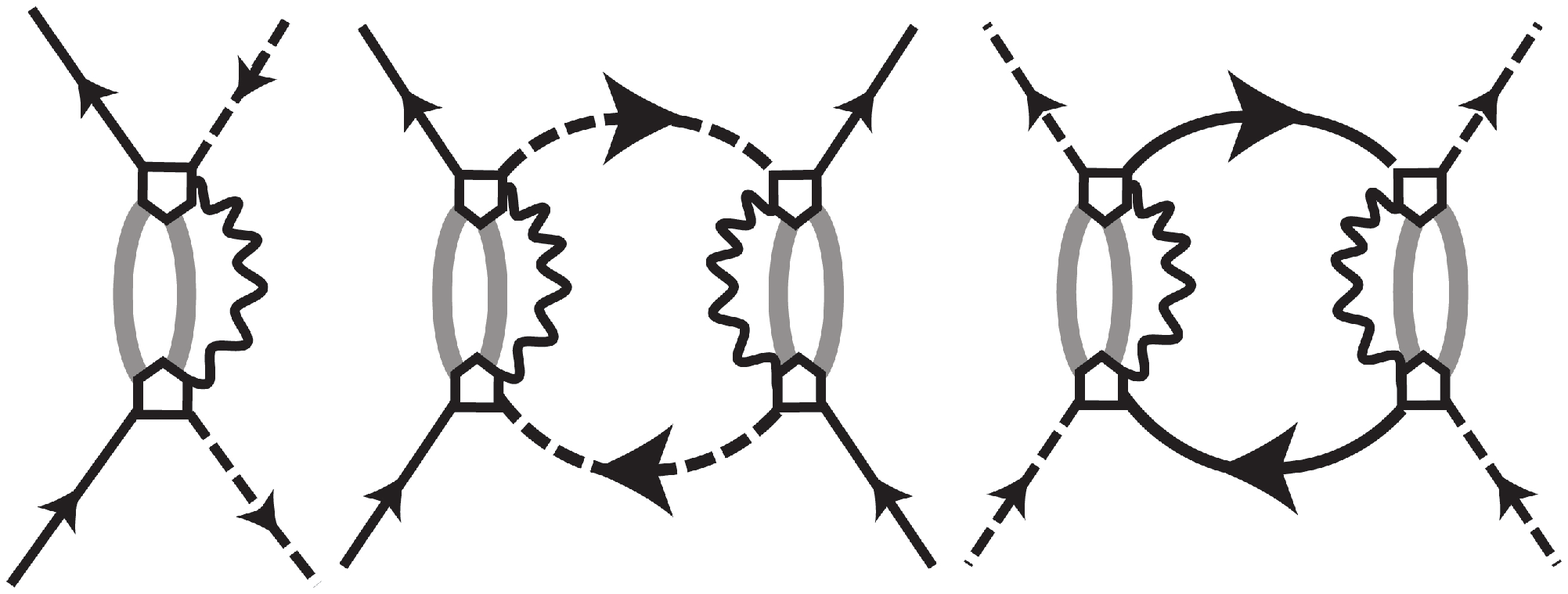}
\caption{Diagram of cofermion-quasiparticle interaction (left), pairing of quasiparticle mediated by cofermion (middle) and pairing of cofermions mediated by quasiparticle (right).  Solid (dashed) lines with arrows are quasiparticle (cofermion) propagators. Wavy lines are charged boson (holon and doublon) propagators and gray bold lines are spin boson propagators.}
\end{flushright}
\label{QPCofermionInteraction}
\end{minipage}
\end{figure}

In the slave boson formalism, an electron is described as a particle composed of a boson (slave boson  representing holon $e$, doublon $d$ and singly occupied up- and down-spin bosons $p_{\uparrow}$ and $p_{\downarrow}$) and a fermion representing the quasiparticle $f_{\sigma}$ with spin $\sigma$ in the form 
$\hatd{c}{i\sigma}=
( \hatd{p}{i\sigma}\hatn{e}{i}
+\hatd{d}{i}\hatn{p}{i\sigma})\hatd{f}{i\sigma}.$
The transfer term $t_{ij} \hatd{c}{i\sigma}\hatn{c}{j\sigma}$ in the original Hubbard model is rewritten in the summation of terms such as 
$t_{ij} \langle\hatd{p}{i\sigma}\hatn{p}{j\sigma}\rangle(\hatd{\Upsilon}{i\sigma}-\hatd{C}{i\sigma})(\hatn{\Upsilon}{i\sigma}-\hatn{C}{i\sigma})$,
where $\hatd{C}{i\sigma}=\hatn{e}{i}\hatd{f}{i\sigma}$. A Stratonovich-Hubbard field $\Upsilon$ introduced here plays the role of the cofermion.
Beyond the previous theories formulated to consider only the mean-field or its Gaussian fluctuations of the slave boson condensate\cite{slavebosonKotliar,Castellani}, the present formalism thus naturally takes into account collective excitations of slave bosons and quasiparticles as cofermions\cite{YamajiPRL,Yamaji2011}.  The above rewritten transfer contains the hybridization of a quasiparticle $f$ with a cofermion $\Upsilon$ after averaging out fluctuating charge bosons and spin bosons as $\langle\hatd{p}{i\sigma}\hatn{p}{j\sigma}\rangle$ in the decoupling approximation.  
The hybridization is scaled by $t_{ij}$ and represents the physical process that the cofermion is dissolved to and recombined from a quasiparticle and a holon (or doublon)
in binding-unbinding fluctuations.
Thus the effective Hamiltonian is 
\begin{equation}
H = \sum_{k} \xi(k) \hatd{f}{k}\hatn{f}{k} + \sum_{k} \alpha(k) \hatd{\Upsilon}{k}\hatn{\Upsilon}{k} + \sum_{k}\Delta(k)\hatd{f}{k}\hatn{\Upsilon}{k}+{\rm H.c.},
\label{QP-CF}
\end{equation}
where the cofermion dispersion $\alpha(k)$ is determined by $t_{cf}$, while the quasiparticle dispersion $\xi(k)$ is also a renormalized one in the absence of the cofermions. 
In the cofermion theory\cite{YamajiPRL}, the obtained hybridization amplitude $\Delta(k)$ is illustrated in Fig.\ref{Delta(k)} for a representative parameter value. 

From Eq.(\ref{QP-CF}), the quasiparticle dispersion given by
\begin{equation}
\Xi(k) = \frac{\xi(k)+\alpha(k)}{2}\pm \frac{\sqrt{[\xi(k)-\alpha(k)/\gamma(k)]^2 + 4\Delta(k)^2/\gamma(k)}}{2}
\label{Xi}
\end{equation}
 develops a gap structure, where the cofermion renormalization factor $\gamma(k)^{-1}$ is also considered.  A remarkable feature is that $\Delta(k)$ is negative definite over the curve $\xi(k)-\alpha(k)/\gamma(k)=0$, leading to the $s$-wave-like full gap and has the amplitude scaled by the transfer. We regard the gap $-2\Delta(k)/\sqrt{\gamma(k)}$ along $\xi(k)-\alpha(k)/\gamma(k)=0$ shown in Fig.\ref{Delta(k)}(b) as the pseudogap. The zero surface associated with the pseudogap is given by the cofermion pole, $\omega=\alpha(k)/\gamma(k)$. Here the incoherent parts of the quasiparticle and cofermion are ignored.  
This excitonic gap is entirely different from the Mott gap precursor and from the superconducting gap as well.  In fact, the pseudogap requires the underlying remnant of the Mott gap to exist but is generated as a lower-energy hierarchy developed with the cofermion.

The cofermion theory reproduces a number of puzzling experimental results as well as remarkable numerical results obtained by the cDMFT and quantum Monte Carlo simulations. For instance, in addition to the consistent doping dependence of the pseudogap with the full-gap structure, the whole structure of the spectral weight $A(k,\omega)$ with upper and lower Hubbard bands and the quasiparticle dispersion, the hole pocket/arc structure of the Fermi surface at $\omega=0$, asymmetric density of states in terms of the $\omega>0$ and $\omega<0$ parts, and growth of the density of states with increasing doping concentration, indicated in the specific heat, are very consistent with the experimental and cDMFT indications\cite{YamajiPRL}.

\section{Superconductivity}\label{Sec4}

The newly established cofermion offers a novel mechanism of superconductivity\cite{Yamaji2011}.
When the charged bosons, representing the holon or doulon degrees of freedom in the slave boson formalism, mediate the pairing, the pairing occurs not only in the two quasiparticles but also between a cofermion and a quasiparticle.  The latter channel has first been considered by two of the authors as what reinforces and enhances the pairing gap, because the mixture of the pairings between the quasiparticles and the cofermions works cooperatively. The mediation of the pairing by the charged boson is even further cooperatively enhanced by adding the spin boson exchange that is basically the same as the spin fluctuation mediated pairing.  

The involvement of the cofermion gives a strong insight to solve the puzzle of the dichotomy known in the cuprates. The dichotomy is experimentally identified by the growth of the pseudogap coexisting with the $d$ wave gap in the superconducting phase both mainly in the antinodal region\cite{InhomogeneousARPES}. Since the pseudogap in the antinodal region appears to enhance the insulating behavior rather than the pairing precursor, it appears to severely compete with the growth of the superconducting $d$-wave gap in this antinodal region.  The puzzle is why such a $d$-wave gap having the largest gap in the antinodal region develops under the insulating pseudogap in the same region.  

Our numerical and theoretical findings solve this puzzle: First, as we have clarified, the pseudogap does not have the $d$-wave shape but has the $s$-wave-type full gap structure, although the gap amplitude is quantitatively larger in the antinodal region. Therefore, it is roughly neutral between the $d$- and $s$-wave symmetries and is not particularly harmful to the $d$-wave superconductivity.  

Second, the larger pseudogap in the antinodal region is particularly helpful for pairing if the cofermion is involved, because the gap in the antinodal region contains the zero surface of the quasiparticle that is nothing but the cofermion poles.  The cofermion pairing is very efficiently formed, which is a component of the total pairing.  In the physical picture, the cofermion may be regarded as the incoherent side of the quasiparticle trapped by holons.  The pairing of the incoherent particles cooperatively reinforces the pairing of the coherent quasiparticles. 

Spatially inhomogeneous gap has been found in the scanning tunnel spectroscopy(STS)\cite{InhomogeneousSTS}.  Two distinct energy scales of the gap have also been pointed out in ARPES\cite{InhomogeneousARPES}. Since the pairing mechanism proposed here has a dual character of coherent and incoherent contributions, respectively from quasiparticles and cofermions, the large gap region observed in STS and presumably trapped near the dopant position is well understood by the localized cofermion, where the hole trapped near the dopant may make a stronger bound state with the quasiparticle, thus contributing to an incoherent and large gap. The large gap in the antinodal region not necessarily correlated with the superconductivity with the $d$-wave form may also be accounted for by the contribution from the incoherent cofermion pairing. The incommensurate charge modulation found in STS\cite{Hanaguri} may also be a manifestation of localized cofermion-quasiparticle pairs or their condensation into the real space.  More detailed analyses along this idea are certainly an intriguing future problem. 

The second novel pairing mechanism originated from the cofermions is that mediated by the cofermion itself as is illustrated in Fig.\ref{QPCofermionInteraction}.  Because of the weakened binding energy, the cofermion has a large polarizability arising from the dipole moment by the holon and electron. Polarizable cofermion fluctuations are a natural source of the glue for the pairing that has a conceptual similarity to the exciton mediated pairing. Internal degrees of freedom of a cofermion has not been explored in detail in the literature, although the cofermion self-energy has to some extent  been considered\cite{Yamaji2011}.  The dipole or excitonic fluctuations is a significant issue left for future studies.
      
\section{Concluding Remark}\label{Sec5}
In this article, we have examined the origin of the pseudogap formation revealed in the cuprate superconductors as well as in the numerical results of the doped 2D Mott insulator and elucidated it from intuitive and physically transparent picture.  The $s$-wave-like full gap found in the cluster extension of the dynamical mean-field theory for the 2D Hubbard model is naturally understood from the cofermion excitation emergent as a natural extension from the exciton known in the Mott insulating phase to the doped systems. The cofermion concept unifies charge and spin fluctuations because it also merges into the spin exchange in the Mott insulator. The hybridization of the cofermion with the quasiparticle (or in other words, cofermion-quasiparticle scattering) generates a gap identified as the pseudogap.  The cofermion is also the origin of the pocket/arc like structure of the Fermi surface.  The cofermion dispersion hidden in the pseudogap region of $(k,\omega)$ space plays a significant role in realizing unusual metallic properties as well as in the superconducting mechanism. Since the cofermion is charge neutral and is not easy to detect, it is reminiscent of the envisioned dark matter in the universe.  Further detailed studies on the hidden cofermion physics is a challenging issue for future.   


\end{document}